\documentstyle[preprint,prb,aps,graphics]{revtex}

\newcommand{\FIG}[1]{#1}
\newcommand{\PREP}[1]{}

\newcommand{\bfv}{\mbox{\bf v}}
\newcommand{\bft}{\mbox{\bf t}}
\newcommand{\bfe}{\mbox{\bf e}}
\newcommand{\bfn}{\mbox{\bf n}}
\newcommand{\bfV}{\mbox{\bf V}}
\newcommand{\bfk}{\mbox{\bf k}}
\newcommand{\bfx}{\mbox{\bf x}}
\newcommand{\bfB}{\mbox{\bf B}}

\newcommand{\bfE}{\mbox{\bf E}}
\newcommand{\deldelt}[1]{\frac{\partial #1}{\partial t}}
\newcommand{\deldelx}[1]{\frac{\partial #1}{\partial x}}

\begin{document}
\draft
\title{Linear wave propagation in relativistic magnetohydrodynamics}
\author{R. Keppens}
\address{Centre for Plasma-Astrophysics, K.U.Leuven, Celestijnenlaan 200B, 3001 Heverlee, Belgium}
\address{Leuven Mathematical modeling and Computational science Centre, K.U.Leuven, Belgium}
\address{FOM-Institute for Plasma Physics, P.O.~Box 1207, 3430 BE Nieuwegein, The Netherlands} 
\address{Astronomical Institute, Utrecht University, The Netherlands} 
\author{Z. Meliani}
\address{Centre for Plasma-Astrophysics, K.U.Leuven, Celestijnenlaan 200B, 3001 Heverlee, Belgium}
\maketitle

\begin{abstract}
The properties of linear Alfv\'en, slow, and fast magnetoacoustic waves for uniform plasmas in relativistic magnetohydrodynamics (MHD) are discussed, augmenting the
well-known expressions for their phase speeds with knowledge on the group speed. A $3+1$ formalism is purposely adopted to
make direct comparison with the Newtonian MHD limits easier and to stress the graphical representation of their anisotropic linear wave properties using 
the phase and 
group speed diagrams. By drawing these for both the fluid rest frame and for a laboratory Lorentzian frame which sees the plasma move with a three-velocity having an arbitrary orientation with respect to the magnetic field, a graphical view of the relativistic aberration effects is obtained for all
three MHD wave families. Moreover, it is confirmed that the classical Huygens construction relates the phase and group speed diagram in the usual way,
even for the lab frame viewpoint. Since the group speed diagrams correspond to exact solutions for initial conditions corresponding to a localized
point perturbation, their formulae and geometrical construction can serve to benchmark current high-resolution algorithms for numerical
relativistic MHD.
\end{abstract}



\section{Introduction: relativistic MHD}\label{s-intro}
The classical magnetohydrodynamic (MHD) description for the macroscopic dynamics of plasmas offers a uniquely 
powerful, unifying viewpoint on both laboratory and astrophysical plasmas (see e.g. the textbook by Goedbloed and Poedts~\cite{hansbook}). 
The applicability of the MHD description is discussed in many textbooks, and it is
well known that for most laboratory plasmas, the single fluid ideal or resistive MHD model eventually needs 
to be extended towards a multifluid model and/or by including important kinetic effects, since its continuum 
approach to plasma modeling neglected e.g. Landau damping as well as many other velocity-space dependent 
physical phenomena. For many astrophysical plasmas, we face another shortcoming of the MHD model, namely
that it restricts all attention to non-relativistic plasma velocities. This latter shortcoming is easily remedied
by formulating the ideal MHD equations in a frame-invariant relativistic formulation within four-dimensional space-time. The mathematical formulation of the relativistic ideal MHD equations, and their physical 
implications, can be found in (somewhat more technical) monographs by 
Lichnerowicz~\cite{lich} and Anile~\cite{anile}. 
The ideal relativistic MHD (RMHD) equations combine the full set of Maxwell equations with particle and 
tensorial energy-momentum conservation, under the specific assumption that the electric field in the co-moving frame vanishes identically, i.e. $\bfE'=\bf0$, or $\bfE=-\bfv\times\bfB$ as expressed in the familiar 
three-vectors in a fixed laboratory Lorentzian reference frame. The linear wave dynamics in homogeneous relativistic plasmas
are also completely known: just as in classical MHD, a homogeneous magnetized plasma supports slow and fast magnetoacoustic waves, as well as Alfv\'en waves. Expressions for the wave speeds, as obtained for different
reference frames in relative motion to eachother, have been used as basic ingredients for the current suite of relativistic magnetofluid codes~\cite{komis99,balsara,delzanna,leis05,Mign06,giacom2,jcp07,srmhd08,cpc08}. Knowledge of the (fastest) characteristic speeds
embedded in the relativistic MHD equations is essential to obtain timestep stability limits for explicit time stepping strategies, while many modern RMHD codes use more complete information on all linear wave dynamics. 
This information is incorporated in the eigenvectors of the flux Jacobian, with the fluxes appearing in the governing hyperbolic equations for the conserved variables, typically formulated in a 
$3+1$ formulation where a fixed laboratory frame is selected. In this paper, we (re)collect the basic formulae for linear RMHD waves, and present them in a more graphical sense using the classical phase and group speed diagram representations, as originally introduced by Friedrichs. 
The relativistic modifications from the classical MHD results are thereby pointed out, as well as the MHD wave anisotropies.
We give both phase and group speed diagrams for all three RMHD wave families, and explain
how well-known effects for relativistic wave transformations (in particular, relativistic wave aberration)
connect these diagrams when drawn from differing frames in relative motion. The basic connection between phase and group diagrams, by means of a Huygens construction to obtain group from phase speed diagrams, is demonstrated for 
cases where the point perturbation is seen to move past with arbitrary velocity $\bfv$. 

Our results complement the wealth of knowledge on linear and nonlinear (shock) wave properties for relativistic MHD, as collected in 
Anile~\cite{anile} and Lichnerowicz~\cite{lich} and in follow-up studies. An example of the latter is the discussion of the properties of Alfv\'en waves
for both continuous and discontinuous (shock) solutions by Komissarov~\cite{komis97}, emphasizing the wave polarization properties as viewed in different 
Lorentzian reference frames. In a more recent publication~\cite{kalra00}, a dispersion relation for RMHD linear waves in a homogeneous plasma convected at uniform speed was analyzed in the most tractable special cases (aligned velocity and
magnetic field vector). This study adopted a $3+1$ formalism, which is the one adhered to here as well. It is pointed out how this result fits in the 
established theoretical framework, and how it can then be used to generalize the findings for linear waves in relativistic anisotropic MHD formulations~\cite{geb02}, with a pressure difference along and perpendicular to the magnetic field.
In Sect.~\ref{s-hd}, we start our presentation with linear sound wave dynamics in relativistic compressible gases. 
This is then generalized to homogeneous plasmas in Sect.~\ref{s-mhd}. 

\section{Relativistic hydro preliminaries}\label{s-hd}

Since the magnetoacoustic waves are direct generalizations from the gas dynamical sound waves, it is instructive to iterate on the relativistic hydro case first. To obtain the relativistic expression for the sound wave speed, one must perform the usual linearization of the relativistic hydrodynamic equations about an 
equilibrium configuration. We do this for a homogeneous gas only, adopting a strategy which rewrites the 
governing covariant equations in 4-dimensional space-time to equivalent expressions in a $3+1$ formalism.
The $3+1$ formalism splits temporal and spatial derivatives in a fixed Lorentzian reference frame, and shows 
clearly how the equations become more involved with respect to their Galilean versions.
Also, this $3+1$ formalism is the one adopted in modern numerical approaches
as it was already proposed by York~\cite{York79}.
When we choose as primitive variables the entropy $S$, the rest frame proper density $\rho$, and the 
velocity $\bfv$ in our laboratory frame, and temporarily restrict
to a constant polytropic equation of state, where $S=p\rho^{-\gamma}$, we have as governing primitive variable equations
\begin{eqnarray}
\frac{\partial S}{\partial t}  & + & \bfv \cdot \nabla S = 0 \,, \nonumber \\
\frac{\partial \rho}{\partial t} & + & \bfv\cdot \nabla \rho +\frac{\rho h}{u} \nabla \cdot \bfv \nonumber \\
& - & \frac{1}{u \Gamma^2} \bfv \cdot \nabla \left(S \rho^\gamma\right)  = 0 \,,\nonumber \\
\frac{\partial \bfv}{\partial t} & + & \left(\bfv \cdot \nabla\right)\bfv + \frac{c^2}{\rho h \Gamma^2} 
\nabla \left(S\rho^\gamma\right) \nonumber \\
&-& \bfv \left(\nabla\cdot \bfv\right) \left[1-\frac{y c^2}{u}\right] -\bfv \frac{y c^2}{u\rho h \Gamma^2} 
\bfv\cdot\nabla\left(S\rho^\gamma\right)=0 \,.\nonumber
 \\
& & \label{hd}
\end{eqnarray}
In these equations, $c$ is the light speed, and we introduced
\begin{eqnarray}
u =  & h - \frac{v^2}{c^2} \gamma S \rho^{\gamma-1}  & \left[\rightarrow c^2\right] \,, \nonumber \\
y=& \frac{2h}{c^2}-1 -\frac{\gamma^2 S \rho^{\gamma-1}}{c^2(\gamma-1)} & \,\, \left[\rightarrow 1\right] \,,\nonumber \\
h=& c^2+\frac{\gamma S \rho^{\gamma-1}}{\gamma-1} & \left[\rightarrow c^2\right] \,, \label{hd1}
\end{eqnarray}
where the expressions between brackets denote the Galilean limits where the Lorentz factor $\Gamma\rightarrow 1$. 
In this form, no approximations have been made yet
(noting that the generality and the elegance of the covariant formulation
has been lost), and we can clearly identify all terms denoting 
relativistic corrections with respect to the familiar Euler equations. 

The equations~(\ref{hd}) can easily be linearized about a static $\bfv=\bf0$, uniform gas with constant 
entropy and density $S, \rho$. Assuming a plane wave variation $\exp(-i\omega t+i\bfk\cdot\bfx)$ of all 
linear quantities $S_1, \rho_1, \bfv_1$ we arrive at
\begin{eqnarray}
\omega S_1  &=& 0\,,\nonumber \\
\omega \rho_1 &=& \rho \bfk\cdot\bfv_1 \,,\nonumber \\
\omega \bfv_1 &=& \frac{c^2}{\rho h}\bfk\left(S\gamma\rho^{\gamma-1}\rho_1+
\rho^\gamma S_1\right) \,.\label{hd2}
\end{eqnarray}
As in the non-relativistic case, this system admits for 5 solutions, where 3 wave modes are at marginal 
frequency $\omega=0$. These are the entropy wave with arbitrary $S_1$ but without density or velocity 
perturbation, together with the two transverse translations (shear waves).
The physically more interesting modes are compressible perturbations, i.e. sound waves, with 
$\bfk\cdot\bfv_1\ne 0$. These obey the dispersion relation
\begin{equation}
\frac{\omega^2}{k^2 c^2}=\frac{\gamma S \rho^{\gamma-1}}{h}=\frac{\gamma p}{\rho h}=\frac{c_g^2}{c^2} \,. \label{hd3}
\end{equation}
While this expression is specific to a case with a polytropic equation of state, it is also well-known that for the more appropriate Synge equation of state~\cite{synge}, the sound speed $c_g$ is obtained from
\begin{equation}
\frac{c_g^2}{c^2}=\frac{G'}{GzG'+G/z} \,, \label{hd4}
\end{equation}
where $G(z)=K_3(z)/K_2(z)$ denotes the modified Bessel ratio, and $z=\rho c^2/p$. 
The derivative in Eq.~(\ref{hd4}) obeys $G'=G^2-1-5G/z$, due to recurrence relations.
The enthalpy is then found from $\rho h= \rho c^2 G$.
The Synge equation of state ensures that the effective polytropic index changes smoothly from its classical value $5/3$ for a monoatomic gas, to the relativistic value $4/3$ at ultrarelativistic internal energies. In effect, this ensures that the sound speed in relativistic hydro is limited by $c_g\leq c/\sqrt{3}$, in the adiabatic case.

We can also compute the characteristic speeds for relativistic hydro by noting that the Eqs.~(\ref{hd}) allow to
 read off the components of the $5\times 5$ coefficient matrix $W$ in the quasi-linear form (for 1D spatial variation)
\begin{equation}
\deldelt{\bfV} + W \deldelx{\bfV} = 0 \,, \label{hdql}
\end{equation}
where the primitive variables are collected in $\bfV=(S,\rho,v_x,v_y,v_z)^T$. 
When we compute the 5 eigenvalues $\lambda$ of the $W$ matrix, one obtains the characteristic equation
\begin{equation}
\left(\lambda-v_x\right)^3 \left(\lambda^2-2\lambda v_x \frac{1-\frac{c_g^2}{c^2}}{1-\frac{v^2c_g^2}{c^4}}+
\frac{v_x^2\left(1-\frac{c_g^2}{c^2}\right)-c_g^2\left(1-\frac{v^2}{c^2}\right)}{1-\frac{v^2c_g^2}{c^4}}
\right) = 0 \,. \label{hdchar}
\end{equation}
Hence, the characteristic speeds either take the value $\lambda=v_x$ 
(which obviously correspond to the entropy and shear waves from above), while the sound waves are this time 
found from a seemingly more complicated quadratic expression. Naturally, both approaches must
 agree. The key observation is that by computing the characteristic speeds from the $W$ matrix,
we in fact linearized the equations about a moving plasma, and we need to consider how plane waves in 
the gas rest frame transform relativistically to a moving reference frame.
This obviously involves a Lorentz transformation, which for a frame $L'$ with coordinates $(ct',\bfx')$ moving with velocity $\bfv$ with respect to $L$ is given by
\begin{equation} 
\cases{\displaystyle \ \bfx' = \bfx + {\Gamma -1 \over v^2} \,\bfv \,\bfv \cdot \bfx - 
\Gamma \bfv \,t \,, &\cr &\cr \displaystyle \ t' = \Gamma \,\Big(t - {1 \over c^2} \bfv \cdot \bfx \Big) \,, 
\qquad \Gamma \equiv {1 \over \sqrt{1 - v^2 /c^2}} \,.&\cr} \label{lor}
\end{equation}
When we momentarily assume that in $L'$, we have a plane wave with variation $\exp(-i\omega' t' +i \bfk'\cdot\bfx')$, one finds
directly that frame $L$ will still observe a plane wave with variation $\exp(-i\omega t + i \bfk\cdot\bfx)$
with frequency and wavevector given by
\begin{eqnarray}
\omega &=&\Gamma\left(\omega'+\bfk'\cdot\bfv\right) \,,\nonumber\\[2mm]
\bfk &=& \bfk'+\bfv\left[\frac{\omega'\Gamma}{c^2}+\left(\bfk'\cdot\bfv\right)\frac{\Gamma-1}{v^2}\right] 
\,. \label{dop}
\end{eqnarray}
These familiar expressions quantify the relativistic Doppler effect (i.e.\ the change in frequency) and shows that the wave vector is observed to change direction when viewed from a moving vantage point. The
latter effect is known as relativistic wave aberration. 
The inverse formulae (identical with $\omega'\leftrightarrow\omega$, $\bfk'\leftrightarrow\bfk$ and a sign 
change for $\bfv\leftrightarrow-\bfv$) allow to find the phase speed for frame $L$ given 
by $v_{\rm ph}=\omega/k$ from the formula
\begin{equation}
\frac{v_{\rm ph}'^2}{c^2}=\frac{\Gamma^2 \left(v_{\rm ph}-\bfn\cdot\bfv\right)^2}{c^2+\Gamma^2\left(v_{\rm ph}
-\bfn\cdot\bfv\right)^2-v_{\rm ph}^2} \,. \label{phase}
\end{equation}
We introduced the unit vector $\bfn=\bfk/k$. When we invert this formula to find the phase speed 
in frame $L$, we get the quartic from Eq.~(\ref{hdchar}), in which $\lambda\leftrightarrow v_{\rm ph}$ and $v_x\leftrightarrow \bfn\cdot\bfv$.

\begin{figure}
\begin{center}
\FIG{
\resizebox{\textwidth}{!}
{\includegraphics{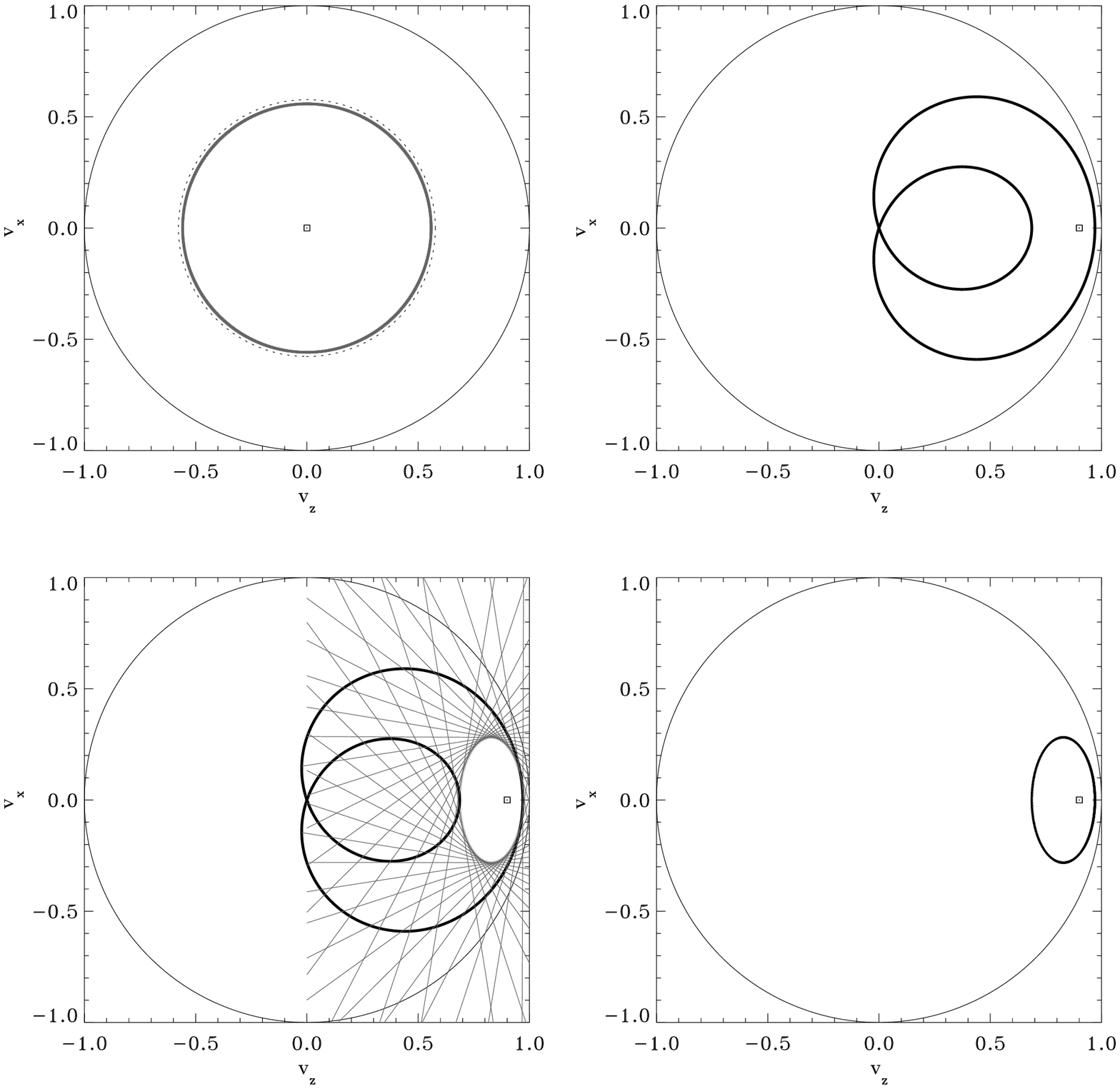}}
}
\end{center}
\caption{}
\label{fighd}
\end{figure}

To complete this discussion on plane wave propagation in uniform gases for special relativity, we now draw 
the phase and group speed diagrams in different reference frames. This is done in Fig.~\ref{fighd}, 
where the $z-x$ plane is drawn. We show at top left the phase $v_{\rm ph}\bfn/c$ and 
group speed $\bfv_{\rm gr}=\partial \omega/\partial\bfk$ in the gas rest frame, 
where we find isotropic sound wave propagation at speed $c_g$ in all directions, 
and where phase and group speed diagrams overlap. We took $\rho=1=p$ and adopted a Synge equation of state.
The sound wave diagram will always be interior to the inner dashed circle which corresponds to the upper 
limit $c/\sqrt{3}$, while the outer circle indicates the light limit.
In the right panel, the phase speed diagram is plotted as seen in a reference frame where the plane wave 
emitter is moving at velocity $\bfv=0.9 c\bfe_z$ along the horizontal $z$-axis. The wave aberration effects 
deformed the single circle to a kind of double spiral form. In the bottom panels, we indicate how a Huygens 
construction then yields the corresponding group speed diagram in the same reference frame. The group 
diagram is what evolves from a point perturbation in a finite time. The final panel just shows this group 
speed diagram in that frame, and demonstrates how the wave front gets ``beamed'' into an asymmetric 
(about the position of the point source) oval shape. We conveniently suppressed the third direction from these diagrams, which would involve revolving the shapes about the $z$-axis.
To conclude this section on sound waves, we note that the group diagram as appropriate for a point perturbation which is seen to pass by at velocity $\bfv=v\bfe_z$ can be obtained in various ways. We saw that graphically, it is found from a Huygens construction. Secondly, we can note that in frame $L'$, the wave fronts are
given by the expressions $x'=v'_{\rm gr,x}(\theta')t'$ and $z'=v'_{\rm gr,z}(\theta')t'$, where the group velocity components are parametrized by the polar angle $\theta'$ in this frame. When we transform these expressions to the lab frame $L$,
we find simply that the wave fronts are found from
\begin{eqnarray}
z & = & \frac{v+v'_{\rm gr,z}}{1+\frac{v v'_{\rm gr,z}}{c^2}} t \,,\nonumber \\
x & = & \frac{v'_{\rm gr,x}}{\Gamma \left(1+\frac{v v'_{\rm gr,z}}{c^2}\right)} t \,, \label{grhd1}
\end{eqnarray}
which is exactly as expected from relativistic velocity addition formulae.
Completely analogous, we may actually work with the relativistic addition rule generally given by
\begin{equation}
\frac{\bfv_{\rm gr}}{c}=\frac{\frac{\bfv}{c}\left(\Gamma+\left(\Gamma-1\right)\frac{\bfv\cdot\bfv'_{\rm gr}}{v^2}\right)+\frac{\bfv'_{\rm gr}}{c}}{\Gamma\left(1+\frac{\bfv\cdot\bfv'_{\rm gr}}{c^2}\right)} \,,\label{add}
\end{equation}
where $\bfv'_{\rm gr}=c_g \bfn'$. Using Eq.~(\ref{dop}) quantifying the relativistic aberration, we then find
\begin{equation}
\frac{\bfv_{\rm gr}}{c}=\frac{\bfn\left(1-\frac{v^2 c_g^2}{c^4}\right)+\frac{\bfv}{c}\left(1\mp \frac{c_g^2}{c^2}\right)\left(\frac{\Gamma^2 c^2}{c_g^2}\Omega \mp \frac{\bfn\cdot\bfv}{c}\right)}
{\frac{\Gamma^2c^2\Omega}{c_g^2}\left(1\mp\frac{v^2 c_g^2}{c^4}\right)+\frac{\bfn\cdot\bfv}{c}\left(1\mp 1\right)}\,. \label{grhd2}
\end{equation}
In this expression, the factor $\Omega$ is given by
\begin{equation}
\Omega=\frac{c_g}{c\Gamma}\sqrt{1-\frac{v^2 c_g^2}{c^4}-\frac{\left(\bfn\cdot\bfv\right)^2}{c^2}\left(1-\frac{c_g^2}{c^2}\right)} \,.\label{ome}
\end{equation}
The two sign combinations in Eq.~(\ref{grhd2}) correspond to parametrizing the wave front using either
the forward, or the backward traveling wave, which is then consecutively plotted in all $2\pi$ directions. The
wave front for the group velocity plotted in Fig.~\ref{fighd} can be checked to agree with all three (differing) parametrizations: the one given in Eq.~(\ref{grhd1}) and both sign combinations of Eq.~(\ref{grhd2}).

\section{Phase and group diagrams for waves in homogeneous plasmas}\label{s-mhd}

To obtain the propagation speeds for linear waves in relativistic MHD, the governing conservation laws in 
tensorial form can be linearized in space-time. The results for the characteristic speeds in RMHD are known
and their derivation from the covariant equations can e.g. be found in Lichnerowicz~\cite{lich} or Anile~\cite{anile}. Here, we revisit these results
in a $3+1$ formalism. The algebra involved is then more cumbersome (see e.g. the results from Kalra and Gebretsadkan~\cite{kalra00}), 
even in the case of linearizing about a stationary, homogeneous plasma. This is partly because of the wave aberration effects, which we mentioned already for the relativistic gas dynamic case, and due to the intrinsic anisotropic nature of the MHD wave families.  
For our purposes, it suffices to note that it is possible to write down the equivalent set of equations in 
a $3+1$ formalism for the primitive variables $(S,\rho,\bfv,\bfB)$,  analogous to the Eqs.~(\ref{hd}) 
for gas dynamics. The equation for the entropy is identical for ideal RMHD, while the added equations for the 
magnetic field are familiar from the non-relativistic case:
\begin{equation}
\cases{
\displaystyle \ \nabla \cdot \bfB = 0  \,, 
\cr
\displaystyle \ \frac{\partial \bfB}{\partial t}- \nabla \times (\bfv \times \bfB) =\bf0 \,.} 
\label{mhd}
\end{equation}
The continuity as well as the momentum equation become fairly cumbersome expressions in a $3+1$ split, and  
we only mention what results from them, after linearizing with plane waves $\exp(-i\omega t+ i\bfk\cdot\bfx)$.

\subsection{Rest frame expressions}\label{ss-rest}

The analysis is tractable for the special case of the plasma rest frame, and a very elegant means to obtain the relativistic 
variants of the slow, Alfv\'en and fast wave speeds can e.g. be found in the appendix from Komissarov~\cite{komis99}. 
For the homogeneous plasma rest frame, linearizing and indicating as usual the background (uniform) quantities 
with $S, \rho, \bfB$, and the linear variables with $S_1, \rho_1, \bfB_1, \bfv_1$, we get
\begin{eqnarray}
\omega S_1  &=& 0 \,,\nonumber \\[2mm]
\omega \rho_1 &=& \rho \bfk\cdot\bfv_1 \,,\nonumber \\[2mm]
\omega \bfB_1&=& \bfB (\bfk\cdot\bfv_1)-\bfv_1(\bfk\cdot \bfB) \,, \qquad\bfk\cdot\bfB_1 = 0 \,,\nonumber \\[2mm]
\omega \bfv_1 &=& \frac{c^2}{w}\bfk\left(S\gamma\rho^{\gamma-1}\rho_1+\rho^\gamma S_1\right) +\frac{c^2}{\mu_0 w}\left(\bfk (\bfB\cdot\bfB_1)-\bfB_1(\bfk\cdot\bfB)\right) \nonumber \\[1mm]
&& +\frac{c^2 (\bfk\cdot\bfB)}{\mu_0 w \rho h} \bfB\left(S\gamma\rho^{\gamma-1}\rho_1+\rho^\gamma S_1\right) \,.\label{mhdlin}
\end{eqnarray}
We here again temporarily assumed a polytropic equation of state, where
the specific enthalpy is $h=c^2+\gamma S\rho^{\gamma-1}/(\gamma-1)$,
and introduced the quantity
\begin{equation}
w=\rho h+\frac{B^2}{\mu_0} \,. \label{w0}
\end{equation}
Note that the relativistic invariant magnetic pressure can be evaluated from the rest frame 
value $p_{\rm mag}=B^2/2\mu_0$.
One can directly compare these expressions~(\ref{mhdlin}) with the non-relativistic ones, as found in many textbooks, such as 
Goedbloed and Poedts~\cite{hansbook}. Only the momentum equation yields an extra term 
(the last term is purely relativistic, and the coefficients for the other terms are changed to involve $w$). 
Not surprisingly then, the 7 wave solutions from classical ideal MHD return in slightly modified form. 
The marginal entropy mode is identical, being the solution at $\omega=0$ for which $S_1\ne 0$ only. 
The Alfv\'en waves return in a virtually unmodified form: they represent solutions with $\bfv_1\ne\bf0$ and $\bfB_1\ne\bf0$ while
\begin{equation}
\rho_1=S_1=\bfk\cdot\bfv_1=\bfk\cdot\bfB_1=\bfB\cdot\bfB_1=\bfB\cdot\bfv_1=0 \,, \label{aw1}
\end{equation}
this time given by the dispersion relation
\begin{equation}
\omega^2=c^2\frac{(\bfk\cdot\bfB)^2}{\mu_0 w} \,. \label{aw2}
\end{equation}
They retain their purely field-sampling property familiar from non-relativistic MHD and we can express their phase $\bfv_{\rm ph}$ and group velocity $\bfv_{\rm gr}$ as follows. For that purpose, assuming $\bfn=\bfk/k$ and denoting by $\vartheta$ the angle between $\bfn$ and $\bfB$, we find
\begin{equation}
\frac{\bfv_{\rm ph}}{c} = \frac{B \cos\vartheta}{\sqrt{\mu_0w}}\bfn \,,\qquad 
\frac{\bfv_{\rm gr}}{c}  =  \frac{\bfB}{\sqrt{\mu_0 w}} \,.
\label{aw3}
\end{equation}
The phase diagram consists of two circles with diameter $B/{\sqrt{\mu_0 w}}$, and the group diagram
shows pure pointlike signals along the perturbed fieldline traveling at the Alfv\'en speed, as in the classical case. Unlike the sound speed, which is limited to be less than $c/\sqrt{3}$, the Alfv\'en speed can go up to the light speed (and is equal to it for vacuum conditions).

The compressible modes are obtained from straightforward algebraic manipulations on Eqs.~(\ref{mhdlin}) to the dispersion relation
\begin{equation}
\omega^4 -\omega^2\left[\frac{k^2 c^2}{w}\left(\rho h \frac{c_g^2}{c^2}+\frac{B^2}{\mu_0}\right)+c_g^2\frac{(\bfk\cdot\bfB)^2}{\mu_0 w}\right]+k^2 c^2 c_g^2\frac{(\bfk\cdot\bfB)^2}{\mu_0 w} =0 \,.\label{sf}
\end{equation}
Here we purposely wrote this again in terms of the squared sound speed $c_g^2$, while we can now also introduce the squared Alfv\'en speed $v_A^2=B^2 c^2/\mu_0 w$. The expressions are then generally valid, with the 
expressions for specific enthalpy $h$ and sound speed $c_g$ depending on the equation of state. 
These expressions are also collected in several works, e.g. by Komissarov~\cite{komis99}, in the review on particle acceleration and relativistic shocks by Kirk and Duffy~\cite{kirk99}, and in the early work by Majorana and Anile~\cite{majo87}.
As a result, the phase speeds for slow and fast wave families are found from
\begin{equation}
\frac{\bfv_{\rm ph}}{c} = \frac{v_{\rm ph}}{c}\bfn= \bfn \sqrt{\frac{1}{2}\left(\frac{\rho h}{w}\frac{c_g^2}{c^2}+\frac{v_A^2}{c^2}\right)} \sqrt{1+\delta \cos^2\vartheta \pm a} \,. \label{sf2}
\end{equation}
Here, the symbols $\delta$ and $\sigma$ express the following dimensionless ratios
\begin{equation}
\delta  =  \frac{c_g^2 v_A^2}{\left(\frac{\rho h}{w}c_g^2+v_A^2\right) c^2} \,,\qquad 
\sigma =  \frac{4 c_g^2 v_A^2}{\left(\frac{\rho h}{w}c_g^2+v_A^2\right)^2} \,. \label{sf3}
\end{equation}
The symbol $a$ follows from 
\begin{equation}
a^2=\left(1+\delta\cos^2\vartheta\right)^2-\sigma \cos^2\vartheta \,. \label{sf4}
\end{equation}
Noting that $\rho h/w=1-v_A^2/c^2$, the phase speed for purely parallel propagation reduces to the same expression found in non-relativistic MHD, where we have 
\begin{equation}
\frac{v_{{\rm ph},\parallel}}{c}=\sqrt{\frac{1}{2}\left(\frac{c_g^2}{c^2}+\frac{v_A^2}{c^2}\right) \left[1 \pm \sqrt{1-\frac{4 c_g^2 v_A^2}{(c_g^2+v_A^2)^2}}\right]} \,. \label{sf5}
\end{equation}
The group speed is then written in terms of the orthogonal directions $\bfn=\bfk/k$ and $\bft= [(\bfB/B) \times \bfn]\times \bfn$ as
\begin{equation}
\frac{\bfv_{\rm gr}}{c} = \frac{v_{\rm ph}}{c}\left[\bfn \pm \bft \frac{\left[\sigma\mp 2 \delta\left(a\pm(1+\delta\cos^2\vartheta)\right)\right]\sin\vartheta \cos\vartheta}{2\left(1+\delta\cos^2\vartheta \pm a\right) a}\right] \,. \label{sfgr}
\end{equation}
These can directly be compared with the non-relativistic expressions (see again e.g. Goedbloed and Poedts~\cite{hansbook}) and we can note that all relativistic effects are due to the parameter $\delta$, together with the fact that both sound and Alfv\'en speeds get relativistic corrections. 
As representative examples, we show in Figs.~\ref{figm1}-\ref{figm3} the phase and group diagrams appropriate for the plasma rest frame, for three cases with (again with units making $c=1$)
\begin{itemize}
\item $\rho=1$, $p=1$, $B=1$ for which $c_g=0.558$ and $v_A=0.431$. This is shown in Fig.~\ref{figm1}, and corresponds to a relativistically hot gas with the sound speed approaching its maximal value. The Friedrichs diagrams resemble the classical Newtonian versions.
\item $\rho=1$, $p=1$, and $B=3$ for which $c_g=0.558$ and $v_A=0.820$. This is shown in Fig.~\ref{figm2}, and corresponds to a hot gas with a very strong magnetic field.
\item $\rho=0.01$, $p=0.001$, and $B=1$ for which $c_g=0.349$ and $v_A=0.994$. This case has the Alfv\'en speed nearly reaching the light limit and is for a cold
plasma. Note how both Alfv\'en and slow group diagrams correspond to extreme anisotropic, field-sampling perturbations.
\end{itemize}
All these assume a Synge equation of state. We can note that the phase speeds obey a strict ordering where slow (S), Alfv\'en (A) and fast (F) obey
$0\leq v_{\rm ph,S}\leq v_{\rm ph,A}\leq v_{\rm ph,F} \leq c$ for any direction $\bfn$, and in the plasma rest frame appear
symmetric about the zero value for forward and backward propagating waves. 
The fact that these rest frame Friedrichs diagrams generalize their Newtonian counterparts was already pointed out in the early lecture notes by 
Newcomb~\cite{newcomb}.

\subsection{Lab frame expressions}\label{ss-lab}

To make the RMHD wave dynamics complete, we still need to show phase and group diagrams for inertial frames in relative motion to the homogeneous plasma rest frame. This can indeed be done by linearizing about a moving (i.e. stationary) uniform plasma, as was pursued in Kalra and Gebretsadkan~\cite{kalra00} for the special case where the movement $\bfv$ was aligned with the uniform field $\bfB$, but this is by far the most algebraically complex means to do so. 
It must be realized that the more general result (for arbitrary orientation between $\bfv$ and $\bfB$) was already given
in the textbook by Lichnerowicz~\cite{lich} and Anile~\cite{anile}, obtained from the more appropriate covariant
form (i.e. valid in any reference frame). Still, because much intuition has been built up in classical MHD in a $3+1$ framework, and modern
special relativistic codes exploit this as well, we now discuss the RMHD waves in the lab frame using the $3+1$ split.
In fact, we then just need one additional equation, relating the three-vector magnetic fields in $L'$ (moving away with speed $\bfv$) and $L$, by the well known Lorentz formula (here for ideal RMHD)
\begin{equation}
\bfB'=\frac{\bfB}{\Gamma}+\bfv \frac{\bfv\cdot\bfB}{c^2}\frac{\Gamma}{\Gamma+1} \,. \label{lorb}
\end{equation}
If we indicate the rest frame as $L'$, we know from Eq.~(\ref{aw2}) that 
$(v'_{\rm ph}/c)^2=\left(\bfB'\cdot\bfn'\right)^2/\mu_0 w'$. Then, combining Eq.~(\ref{lorb}), 
the wavevector transformation from Eq.~(\ref{dop}) and
the phase speed relation in Eq.~(\ref{phase}), we obtain the phase speed for Alfv\'en waves in a frame where the
plane wave emitters move away with velocity $\bfv$ as
\begin{equation}
\frac{\bfv_{\rm ph}}{c}=\frac{v_{\rm ph}}{c}\bfn=\left(\frac{\bfn\cdot\bfv}{c}\pm \frac{\bfn\cdot\bfB}{\Gamma^2\left[\sqrt{\mu_0\rho h +2 p_{\rm mag}}\pm \frac{\bfv\cdot\bfB}{c}\right]}\right) \bfn \,. \label{awlab}
\end{equation}
The magnetic pressure can be obtained in the lab frame from 
\begin{equation}
2 p_{\rm mag}=\frac{B^2}{\Gamma^2}+\frac{(\bfv\cdot\bfB)^2}{c^2} \,.\label{pmag}
\end{equation}
It is straightforward to show that Eq.~(\ref{awlab}) gives once more two circles through the origin, and that correspondingly, the group speed diagram still consists of two point perturbations, this time seen to propagate on the
fieldline which was originally through the origin at the time of the initial pointlike perturbation. The group speed in the lab frame can be obtained by 
using the rest frame expression $\bfv'_{\rm gr}/c=\pm\bfB'/\sqrt{\mu_0 w'}$ and use velocity addition rule in Eq.~(\ref{add}) to get
\begin{equation}
\frac{\bfv_{\rm gr}}{c}=\frac{\bfv}{c}\mp \frac{\bfB}{\Gamma^2\left(\sqrt{\mu_0\rho h+2 p_{\rm mag}}\mp\frac{\bfv\cdot\bfB}{c}\right)} \,. \label{awgrlab}
\end{equation}
The phase and group speed diagrams for the Alfv\'en waves, in a case where $\bfB$ is oriented along the
$z$-axis and the source moves away with a velocity $\bfv=0.9[\sin(\pi/4)\bfe_x+\cos(\pi/4)\bfe_z]$ (without loss of generality taken in the $x-z$ plane), 
are shown in Fig.~\ref{figmaw}, for similar values as used in Figs.~\ref{figm1}-\ref{figm3}. 
The diameter of the circles (actually spheres, but we suppressed the third dimension in the plots)
that define the phase diagram is set by the forward and backward group velocity,
and clearly, the part of $\bfv$ perpendicular to $\bfB$ merely 
displaces the field line originally through the origin (in accord with the frozen-in nature of ideal MHD), on which the pointlike Alfv\'en signals propagate.  

Finally, for the slow and fast wave families, the phase speed in the lab frame is again best computed from transforming the rest frame expression in Eq.~(\ref{sf}) to the lab frame. This was already pointed out by Komissarov~\cite{komis99}, and in doing so we obtain the quartic
\begin{eqnarray}
&&\rho h \left(c^2-c_g^2\right)\Gamma^4\left(\frac{v_{\rm ph}}{c}-\frac{\bfn\cdot\bfv}{c}\right)^4-\left(1-\frac{v_{\rm ph}^2}{c^2}\right) \left\{ \Gamma^2\left(\rho h c_g^2+2 p_{\rm mag}c^2\right)\left(\frac{v_{\rm ph}}{c}-\frac{\bfn\cdot\bfv}{c}\right)^2 \right. \nonumber\\[2mm]
&& \left. - c_g^2\left[ \Gamma\left(\frac{\bfv}{c}\cdot \frac{\bfB}{\sqrt{\mu_0}}\right)\left(\frac{v_{\rm ph}}{c}-\frac{\bfn\cdot\bfv}{c}\right) - \frac{\bfn\cdot\bfB}{\Gamma\sqrt{\mu_0}}\right]^2\right\} = 0 \,. \label{sflab}
\end{eqnarray}
This expression has been known for a long time~\cite{lich,majo87} (but in perhaps less transparant notation than the one adhered to here), and is the one used in RMHD codes to compute 
the fastest wave speed from, needed for explicit time integration strategies~\cite{komis99,balsara,delzanna,leis05,cpc08}. 
The cold plasma limit is obtained for $c_g=0$, and has degenerate slow waves. 

We plot in Fig.~\ref{figall} the phase diagrams for the four solutions from the quartic Eq.~(\ref{sflab}), as they are
seen in the lab frame, for the two cases from Figs.~\ref{figm2}-\ref{figm3}. The Alfv\'en phase speeds are shown as well.
We plotted the phase speeds only in the $x-z$ plane (plotted is thus $\bfv_{\rm ph}=v_{\rm ph}\bfn$ by varying $\bfn$ in $x-z$ over $2\pi$), and it should 
be noted that the suppressed third dimension is no longer a mere revolution about some axis as soon as the velocity $\bfv$ is not aligned with $\bfB$.
For the particular case where $\bfv$ is aligned with $\bfB$ (the case in the study by Kalra and Gebretsadkan~\cite{kalra00}), it is merely to be rotated 
about the $z$-axis.
In any case, the visual representation of the phase diagram is arguably the most intuitive way to illustrate the
combined effects of the inherent anisotropic nature of MHD wave propagation (even in uniform media), with the
added complexity brought in by relativistic wave aberration. As in the case of the sound waves, as we vary the
unit vector $\bfn$ over $2\pi$ (or the full $4\pi$ steradian), the pair of forward/backward slow waves will trace out the same curve (surface) in a 
different parametrization, and similarly for the fast wave pair. It is seen from
Fig.~\ref{figall} that the fast wave phase diagram deforms much like the gas dynamic sound wave.
As the transition from the rest frame to the lab frame involves nothing else than what is given by Lorentz transformations, the ordering of the
forward and backward waves remains in any direction, but the symmetry about the zero velocity is obviously lost, and we get
\begin{equation}
 -c \leq v_{\rm ph,F-} \leq v_{\rm ph,A-} \leq v_{\rm ph,S-} \leq v_{\rm ph,S+} \leq v_{\rm ph,A+} \leq v_{\rm ph,F+} \leq c \,. \label{order}
\end{equation}
The two examples drawn in Fig.~\ref{figall} demonstrate that the manner in which the Lorentz transformation deforms the rest frame version is
truly nontrivial, and can give rise to easthetically appealing diagrams.

Finally, we demonstrate in Fig.~\ref{figgrlab} how one can again use a mere 
Huygens construction to obtain the group speed diagram appropriate for a point perturbation that is seen to pass by at velocity $\bfv$. The left panel
only shows the leftward part of the two cusp-shaped slow signals that arise from performing the Huygens construction on the slow phase
curve drawn in the left panel of Fig.~\ref{figall}. The right panel repeats the fast phase diagram, and demonstrates the group diagram which is now
a fully three-dimensional surface (in the figure seen in the $x-z$ cross-section). In the special case with
aligned $\bfv$ and $\bfB$, we also conclude that the transformation from Eq.~(\ref{grhd1}) can be used directly on the
group diagram found in the rest frame. The result of this for the cases from Fig.~\ref{figm1}-\ref{figm3} is shown in Fig.~\ref{figallh}, when
the velocity is at 90\% of the light speed. 
Closed form expressions for the group speed diagram in the lab frame are not given explicitly for arbitrary orientation between $\bfv$ and $\bfB$, 
but the procedure to obtain them is now clear in principle (but an algebraic nightmare): we need to perform relativistic
speed addition using the rest frame expressions from Eq.~(\ref{sfgr}), and thereby transform not just the normal, but also the tangential components of the 
magnetic field to the wave front from frame $L'$ to the lab frame $L$. The
geometric Huygens construction is a much easier means to realize this. 

\section{Conclusions}\label{s-con}

We provided graphical insight into the phase and group speed properties for RMHD waves in uniform, stationary plasmas.
The added value of knowing the group speed diagrams in different laboratory frames is foremost important to test the current suites of RMHD codes: we here
pointed out how to obtain the exact solution (for Alfv\'en in closed form, and for fast and slow, graphically) to the initial value problem where
a pointlike (linear) perturbation is set up in an overall stationary, homogeneous plasma. Varying the governing
equilibrium parameters (plasma beta, looking at the special case where $c_g=v_A$, from warm to cold plasma conditions, from low to high advection speeds), 
one can found out more about intrinsic limitations in the numerical treatment (numerical diffusion, dispersion, stability, etc.). The fact that pointlike 
perturbations in stationary plasmas will give group diagrams that can be very demanding on spatial resolution (due to the wave aberration effects combined with MHD wave anisotropy), is yet 
again a clear indication that RMHD codes
will need some form of grid-adaptivity, to handle complex 3D RMHD astrophysical problems.
The linear wave properties discussed here are also relevant to better appreciate the knowledge of discontinuous (shock) wave solutions allowed
by the RMHD equations~\cite{ht,lich,anile,majo87,appl88,kirk99}, or the linear wave modifications encountered in fluid models for 
relativistic plasmas that invoke anisotropic pressure~\cite{geb02} or do not have sufficient collisionality to justify a RMHD viewpoint~\cite{haz1,haz2}. 
Finally, knowledge of the wave properties in uniform media is indispensible to appreciate the significant
modifications encountered when diagnosing waves and instabilities in non-uniform, relativistic MHD
equilibrium configurations. This would lead to MHD spectroscopy~\cite{apjlet} for accretion disks about compact objects, or for
relativistic jets.

\acknowledgements

We acknowledge financial support from the
Netherlands Organization for Scientific Research, NWO grant 614.000.421, and from the flemish FWO, grant G.0277.08.

\newpage
{\bf Figure Captions:}

\vspace*{0.5cm}
{\bf Figure 1:} Group and phase diagrams for a relativistic hydro case with $\rho=1=p$ and a Synge equation of state. Top left: in the gas rest frame,
where the circular group and phase diagram coincide, and lie within the limit $1/\sqrt{3}c$ (dotted curve). The light limit is indicated as well (taking $c=1$). Top right: the phase diagram from a lab viewpoint which saw the source pass at $\bfv=0.9\bfe_z$. Bottom left: a Huygens construction in the same lab viewpoint, to obtain the group diagram for the lab frame. The source position is indicated by a small square. Bottom right:
the group diagram for the lab frame, giving the wave fronts as they appear from a point perturbation passing by at $\bfv=0.9\bfe_z$.

\vspace*{0.5cm}
{\bf Figure 2:} Phase (left) and group (right) diagram in the plasma rest frame, for a case with $\rho=1=p=B$ and a Synge EOS. In the left panel, the
maximal sound speed limit is indicated as well (dotted curve), and both panels also show the light limit.

\vspace*{0.5cm}
{\bf Figure 3:} Phase (left) and group (right) diagram in the plasma rest frame, for a case with $\rho=1=p$ and $B=3$ and a Synge EOS. In the left panel, the
maximal sound speed limit is indicated as well (dotted curve), and both panels also show the light limit.

\vspace*{0.5cm}
{\bf Figure 4:} Phase (left) and group (right) diagram in the plasma rest frame, for a case with $\rho=0.01$, $p=0.001$ and $B=1$ and a Synge EOS. In the left panel, the
maximal sound speed limit is indicated as well (dotted curve), and both panels also show the light limit.

\begin{figure}
\begin{center}
\FIG{
\resizebox{\textwidth}{!}
{\includegraphics{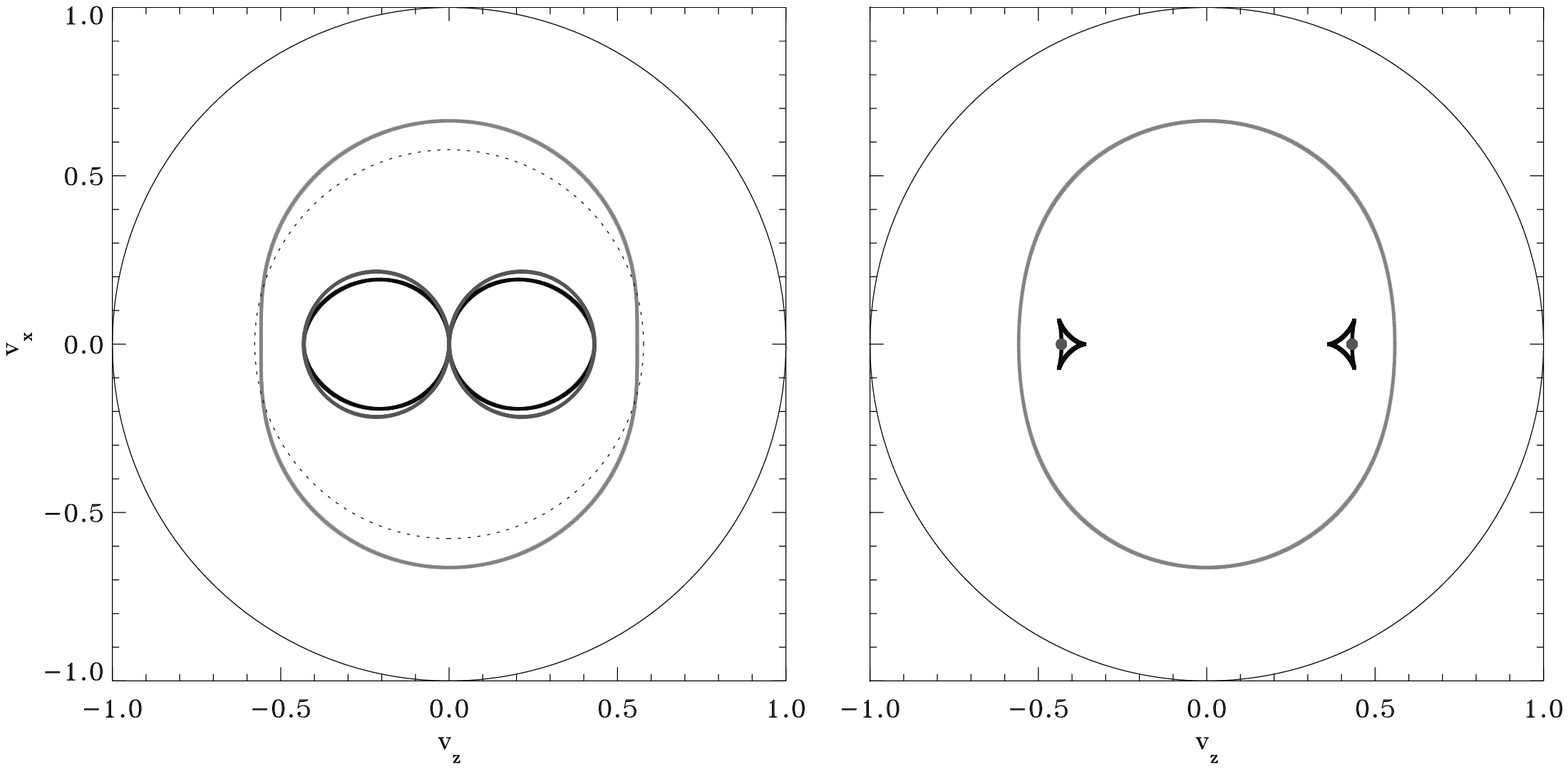}}
}
\end{center}
\caption{}
\label{figm1}
\end{figure}

\vspace*{0.5cm}
{\bf Figure 5:} Phase (left) and group (right) diagram for the Alfv\'en waves only, in the lab frame where the perturbation of the central
(horizontal) field line is seen to move at $\bfv=0.9\left[\sin(\pi/4)\bfe_x+\cos(\pi/4)\bfe_z\right]$. Top panel is for a case where
$\rho=1=p=B$ (as in Fig.~\ref{figm1}), bottom panel for a case where $\rho=0.01$, $p=0.001$ and $B=1$ (as in Fig.~\ref{figm3}).

\begin{figure}
\begin{center}
\FIG{
\resizebox{\textwidth}{!}
{\includegraphics{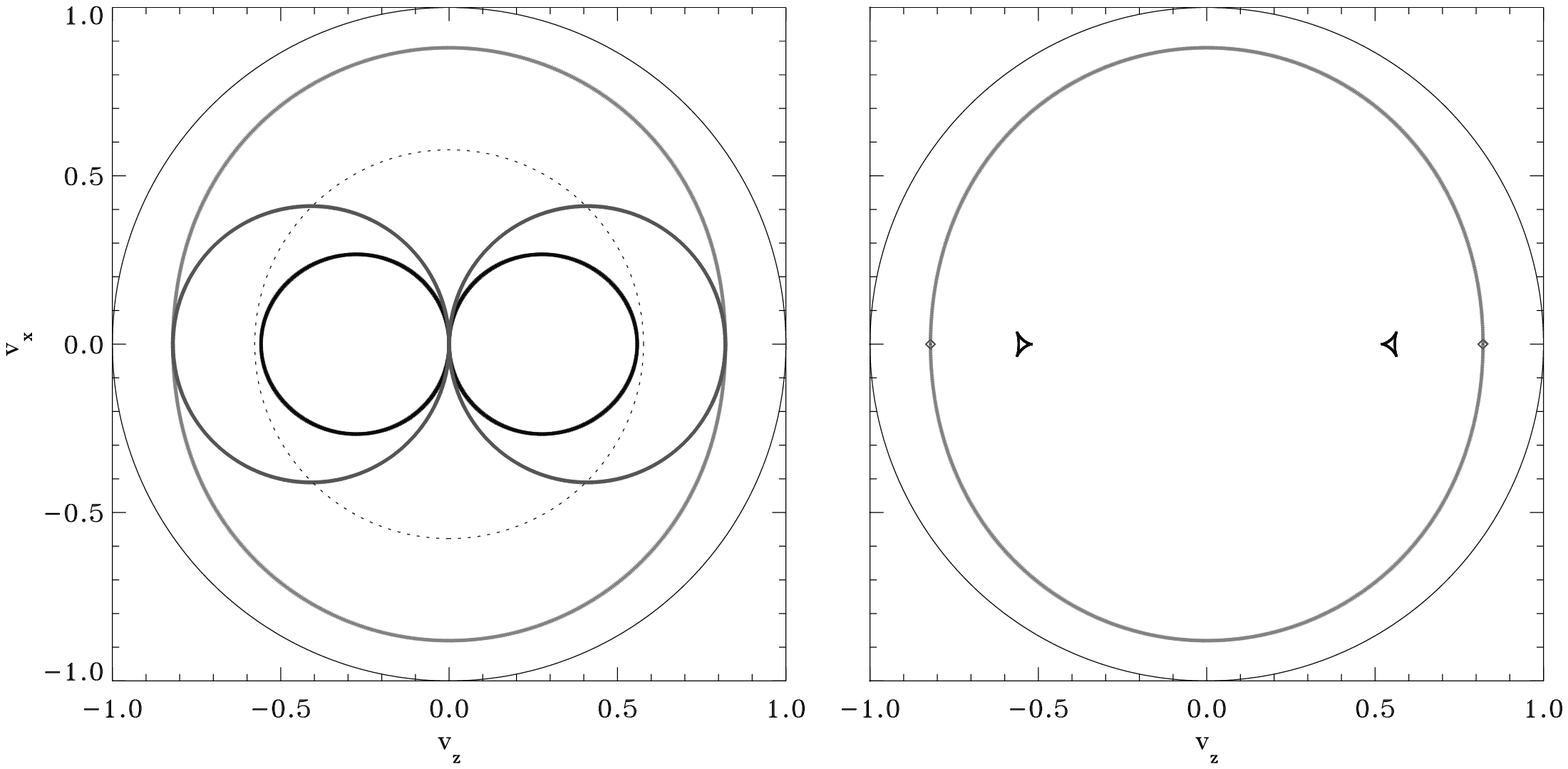}}
}
\end{center}
\caption{}
\label{figm2}
\end{figure}

\begin{figure}
\begin{center}
\FIG{
\resizebox{\textwidth}{!}
{\includegraphics{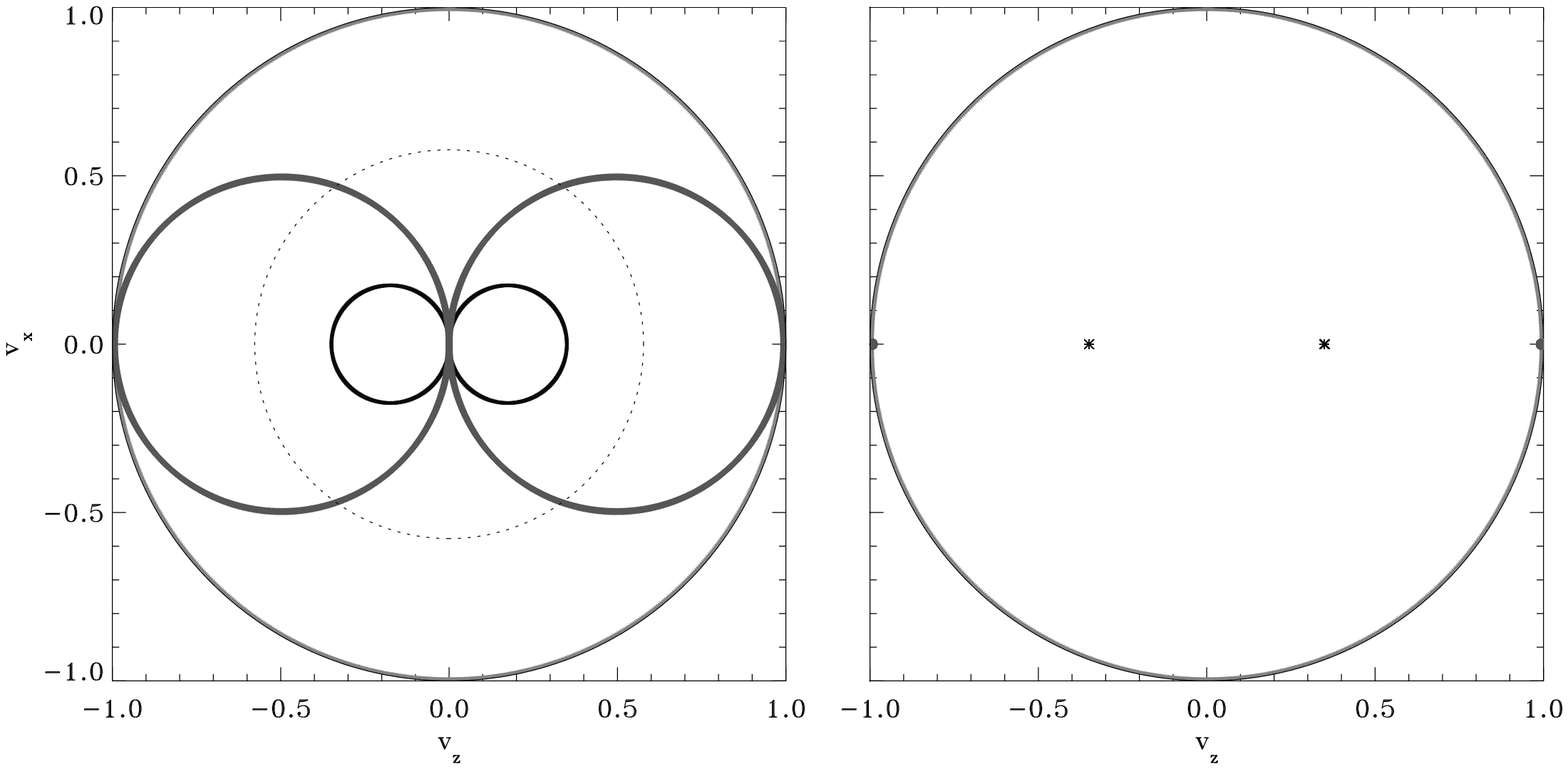}}
}
\end{center}
\caption{}
\label{figm3}
\end{figure}

\begin{figure}
\begin{center}
\FIG{
\resizebox{\textwidth}{!}
{\includegraphics{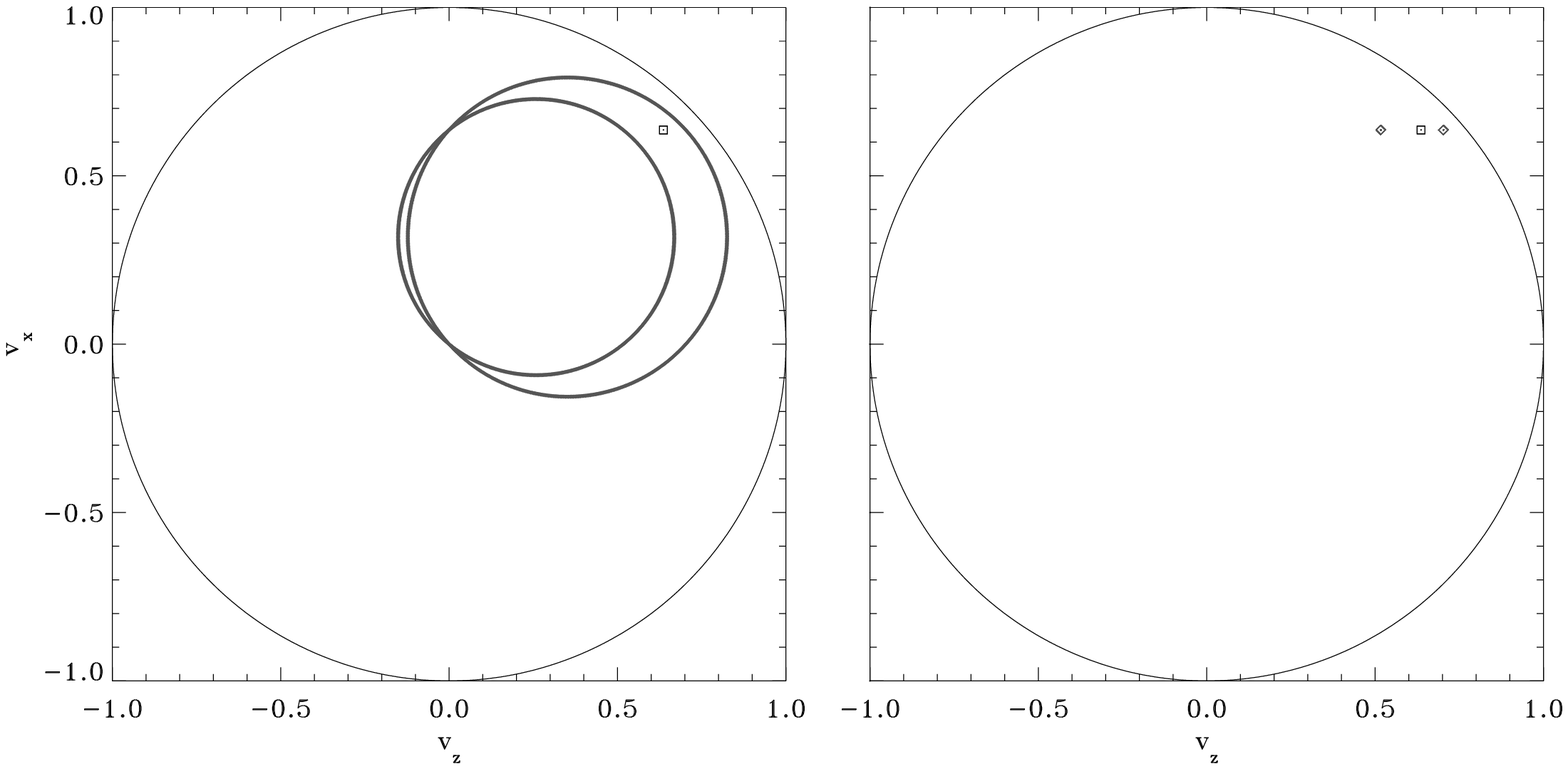}}
\resizebox{\textwidth}{!}
{\includegraphics{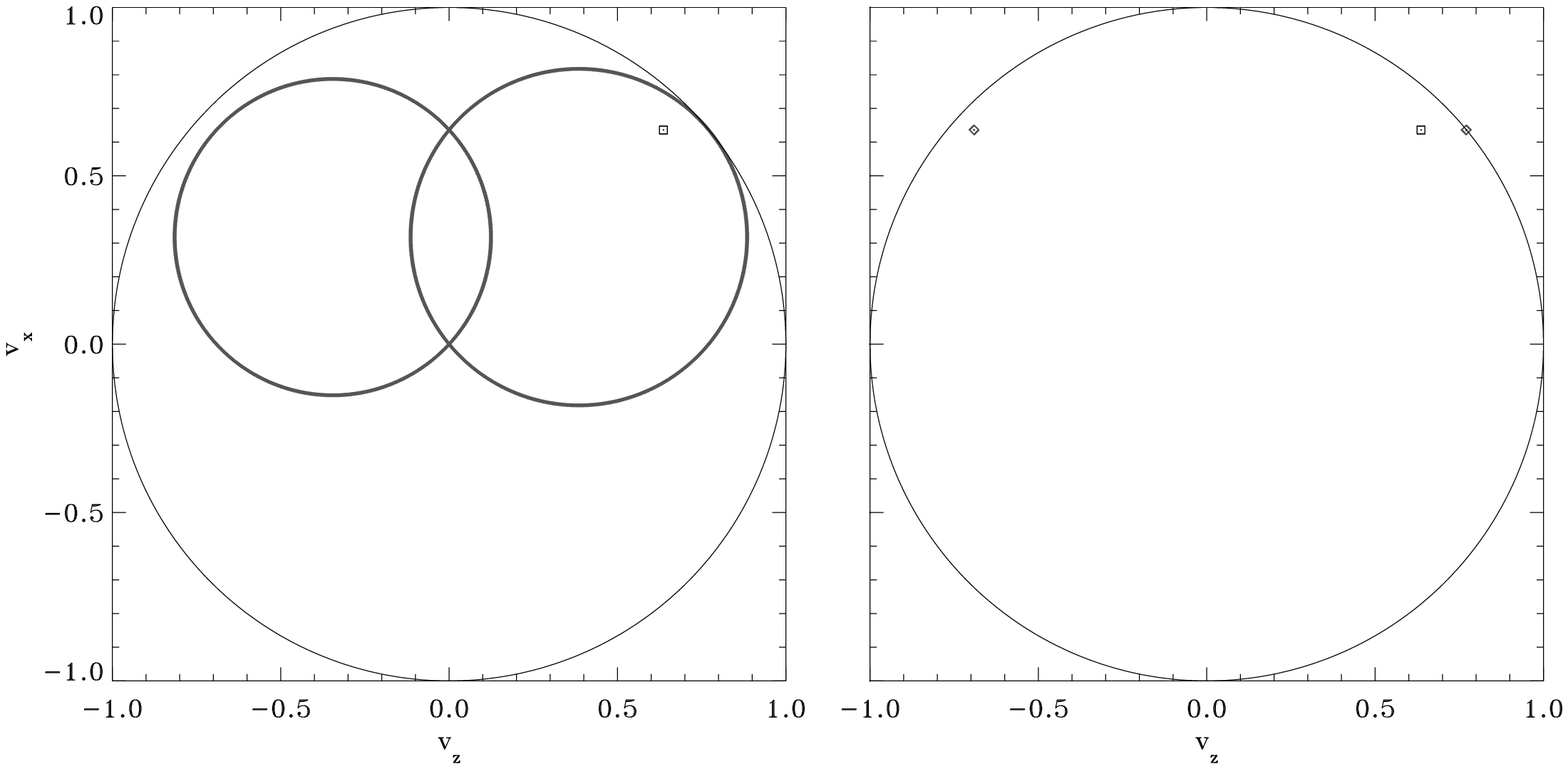}}
}
\end{center}
\caption{}
\label{figmaw}
\end{figure}

\vspace*{0.5cm}
{\bf Figure 6:} Phase diagram for fast, Alfv\'en, and slow waves, in the lab frame where the perturbation of the central
(horizontal) field line is seen to move at $\bfv=0.9\left[\sin(\pi/4)\bfe_x+\cos(\pi/4)\bfe_z\right]$. The case at left is
$\rho=1=p$ and $B=3$ (as in Fig.~\ref{figm2}), at right we have $\rho=0.01$, $p=0.001$ and $B=1$ (as in Fig.~\ref{figm3}).

\begin{figure}
\begin{center}
\FIG{
\resizebox{0.48\textwidth}{!}
{\includegraphics{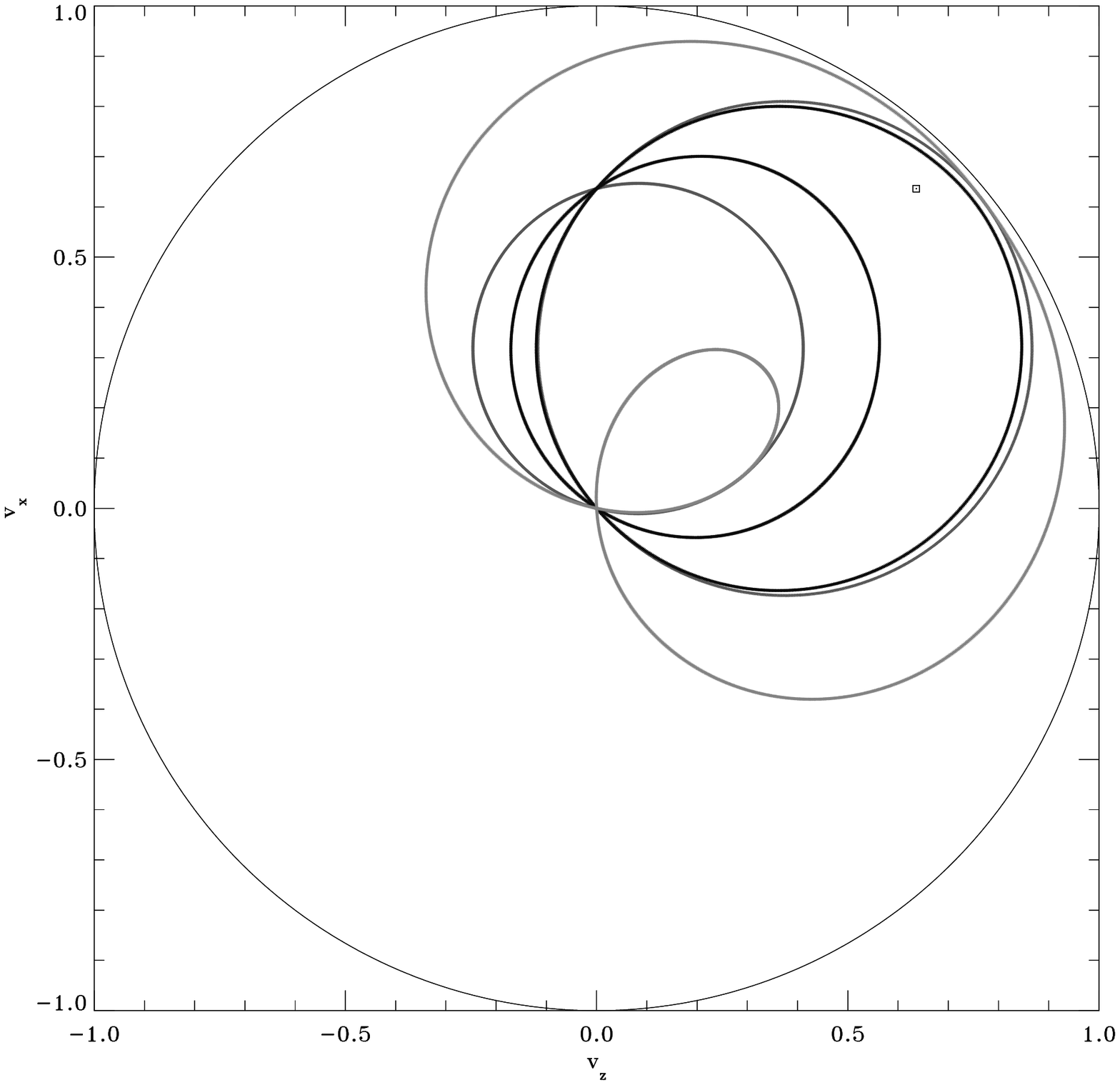}}
\resizebox{0.48\textwidth}{!}
{\includegraphics{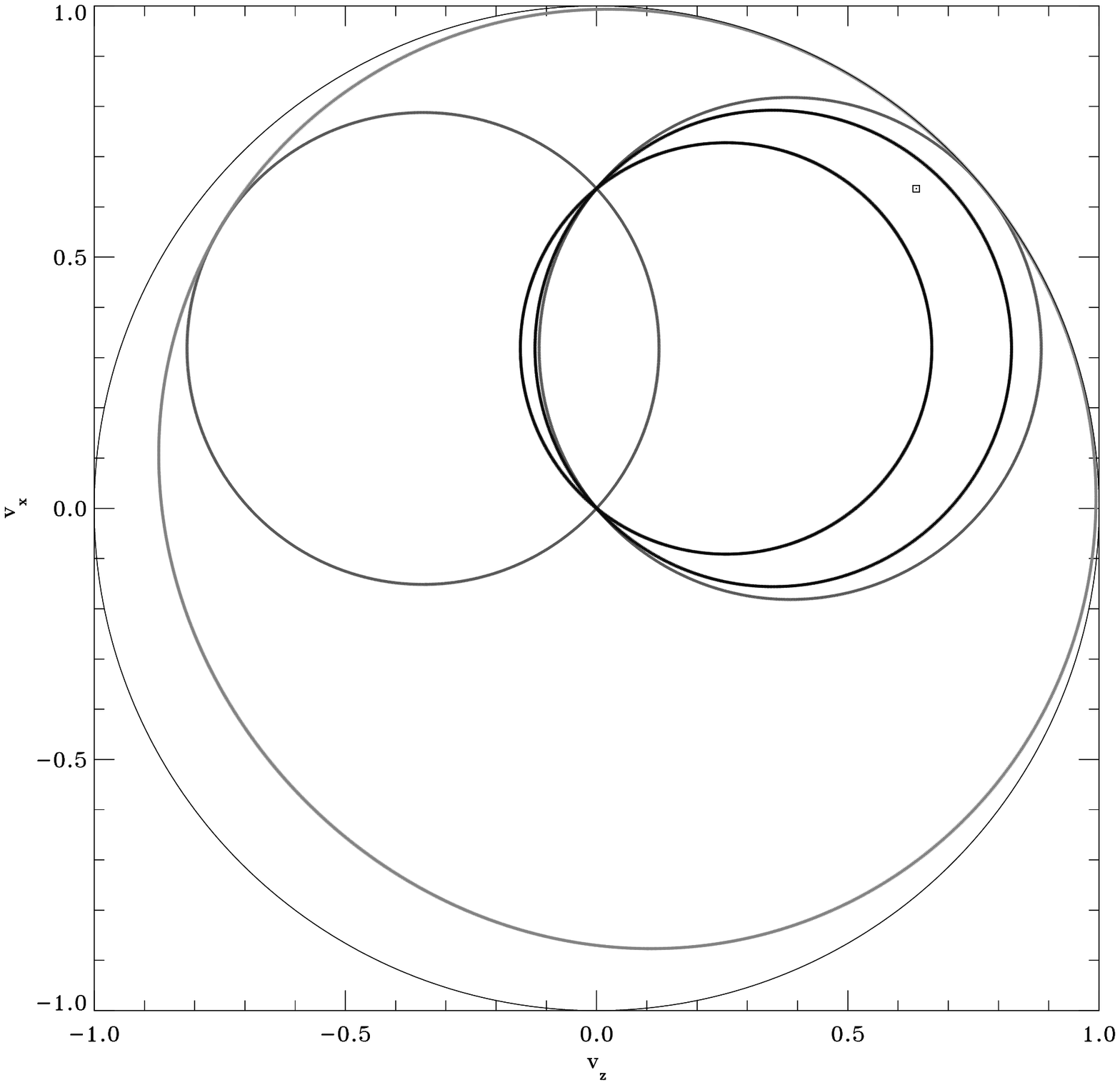}}
}
\end{center}
\caption{}
\label{figall}
\end{figure}

\begin{figure}
\begin{center}
\FIG{
\resizebox{\textwidth}{!}
{\includegraphics{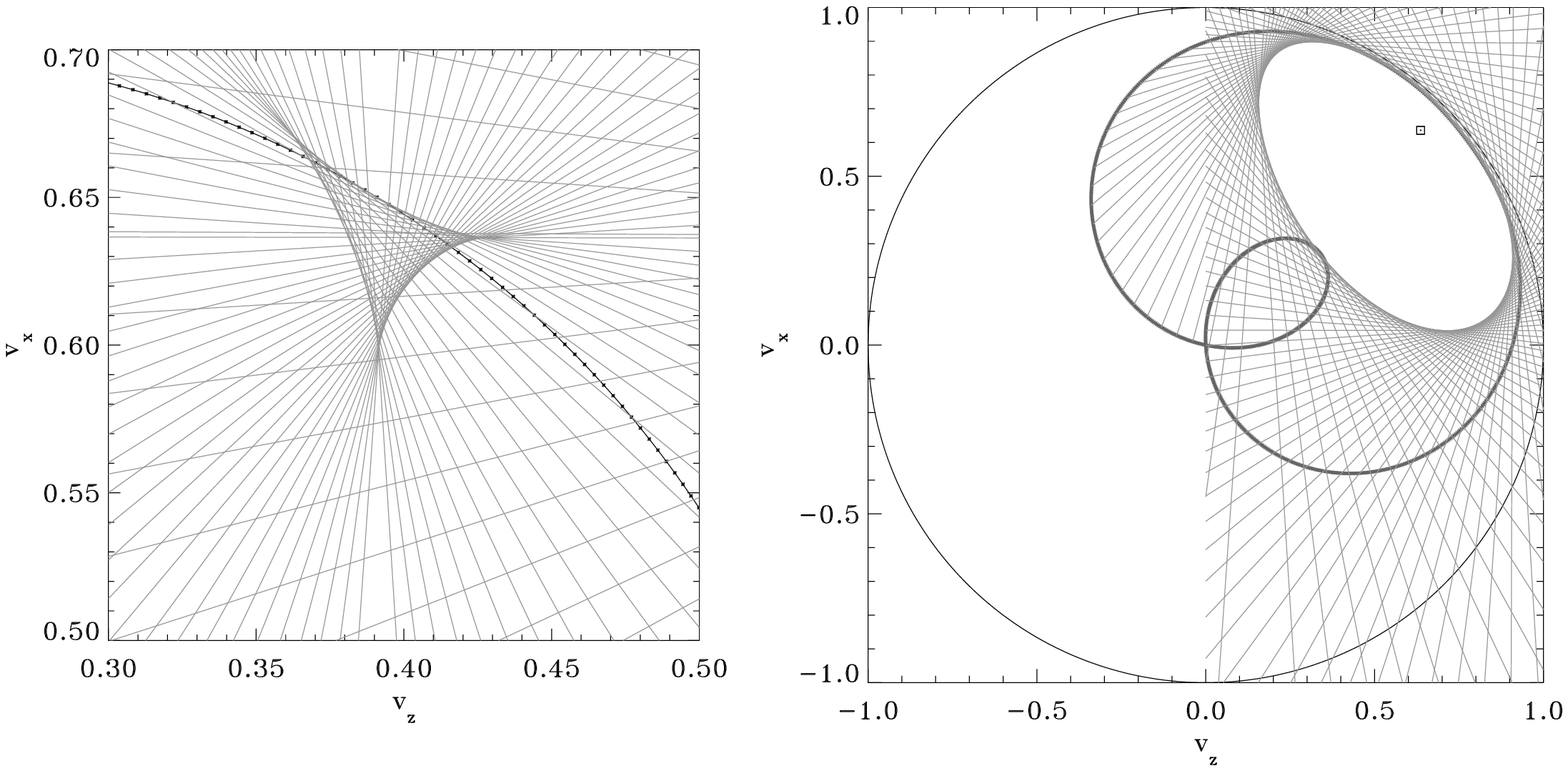}}
}
\end{center}
\caption{}
\label{figgrlab}
\end{figure}

\begin{figure}
\begin{center}
\FIG{
\resizebox{0.48\textwidth}{!}
{\includegraphics{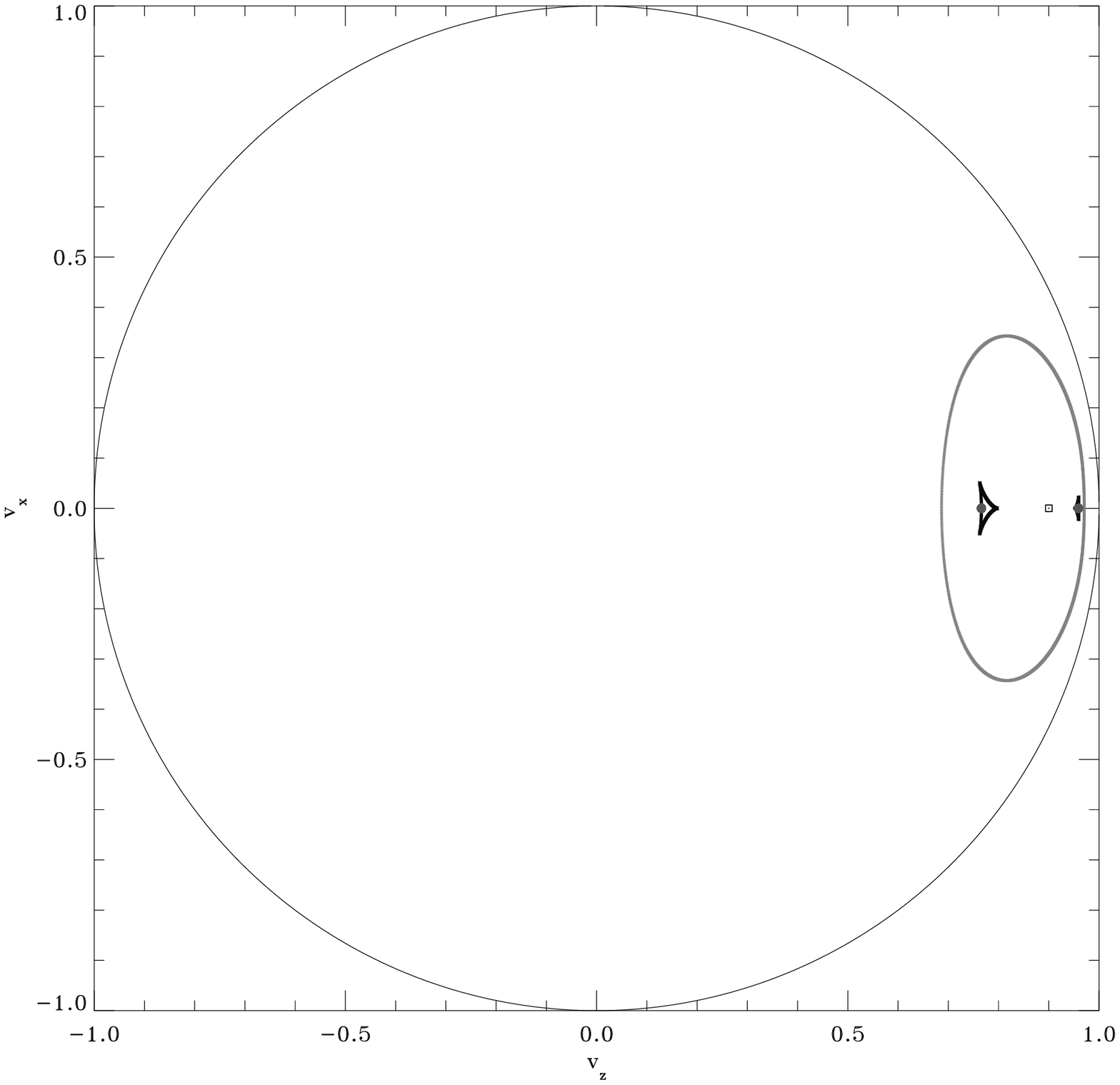}}
\resizebox{0.48\textwidth}{!}
{\includegraphics{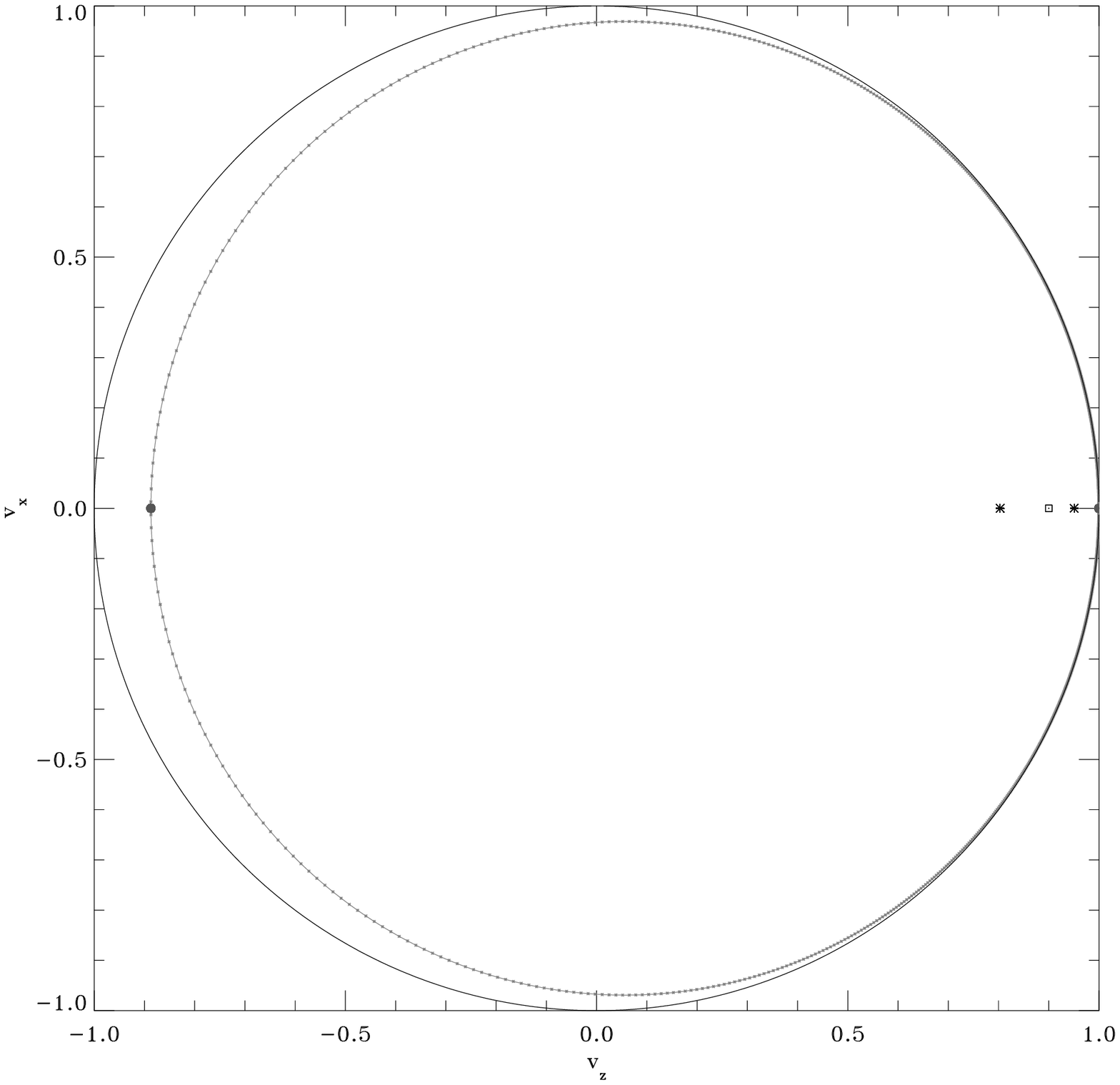}}
}
\end{center}
\caption{}
\label{figallh}
\end{figure}

\vspace*{0.5cm}
{\bf Figure 7:} Group diagram for slow (left) and fast (right) waves, in the lab frame as obtained by means of a Huygens construction.
The perturbation of the central (horizontal) field line is seen to move at $\bfv=0.9\left[\sin(\pi/4)\bfe_x+\cos(\pi/4)\bfe_z\right]$. The case is
$\rho=1=p$ and $B=3$ (as in Fig.~\ref{figm2} and left panel of Fig.~\ref{figall}).

\vspace*{0.5cm}
{\bf Figure 8:} Group diagram for all waves in the lab frame with perturbation seen to move at $\bfv=0.9\bfe_z$. The case is
at left $\rho=1=p=B$ (as in Fig.~\ref{figm1}) and at right $\rho=0.01$, $p=0.001$ and $B=1$ (as in Fig.~\ref{figm3}).

\newpage

\end{document}